\documentclass[10pt,prd,superscriptaddress,amsfonts,amssymb,amsmath,showpacs,twocolumn]{revtex4-2}
\usepackage{bm}
\usepackage{amsfonts}
\usepackage{latexsym}
\usepackage{graphicx}
\usepackage{amsmath}
\usepackage{palatino}
\usepackage{mathpazo}
\usepackage{textcomp}
\usepackage{float}
\usepackage{booktabs}
\usepackage{dcolumn}
\usepackage{booktabs}
\usepackage{multirow}
\usepackage{hyperref}
\usepackage{orcidlink}
\hypersetup{colorlinks,citecolor=blue}
\usepackage{xcolor}

\usepackage{caption}
\usepackage{subcaption}

\begin{document}
	
	\color{black}       

	\title{Dark energy behavior from static equation of state in non-minimally coupled gravity with scalar deformation}
	\author{ N. Myrzakulov
	}
	\email{nmyrzakulov@gmail.com}
	\affiliation{L N Gumilyov Eurasian National University, Astana 010008, Kazakhstan}
	\author{ S. H. Shekh}
	\email{ da\_salim@rediff.com}
	\affiliation{L N Gumilyov Eurasian National University, Astana 010008, Kazakhstan}
	\affiliation{Department of Mathematics, S.P.M. Science and Gilani Arts, Commerce College, Ghatanji, Yavatmal, \\Maharashtra-445301, India.}

\author{S. R. Bhoyar}
\email{drsrb2014@gmail.com}
\affiliation{Department of Mathematics, Phulsing Naik Mahavidyalaya Pusad -445216 Distt.Yavatmal (M.S) India}
	
	\author{Anirudh Pradhan}
	\email{pradhan.anirudh@gmail.com}
	\affiliation{Centre for Cosmology, Astrophysics and Space Science, GLA University, Mathura-281 406,\\ Uttar Pradesh, India}
	
	\begin{abstract}
		\textbf{Abstract:} 
			In this study, we explore the cosmological implications of a modified gravity theory characterized by the function \( f(R, \Sigma, T) \), where \( R \) is the Ricci scalar, \( \Sigma \) represents a geometric deformation term, and \( T \) denotes the trace of the energy-momentum tensor. The model is reconstructed under the framework of three fundamental energy conditions: Null Energy Condition (NEC), Dominant Energy Condition (DEC), and Strong Energy Condition (SEC). We derive the corresponding Hubble parameter \( H(z) \) for each case and constrain the free parameters \( H_0 \), \( \alpha \), and \( \beta \) using the latest Cosmic Chronometer (CC) and Pantheon+ Type Ia Supernova datasets. A thorough analysis of physical quantities such as pressure, energy density, and the equation of state parameter is carried out. Furthermore, diagnostic tools including the deceleration parameter \( q(z) \), statefinder parameters \( \{r, s\} \), and the $O_m(z)$ diagnostic are employed to assess the models viability. Our findings suggest that the model can consistently describe late-time cosmic acceleration and offers distinct behavior under different energy conditions, all while aligning with current observational data.\\

	\end{abstract}

	\maketitle
	\section{Introduction}
	
The discovery of the universe's accelerated expansion has been corroborated by multiple independent observational studies \cite{Perlmutter1997, Riess2007, Spergel2003, Hawkins2003, Eisenstein2005}. However, despite its success in explaining a wide range of gravitational phenomena, \textit{General Relativity} (GR) alone does not provide a satisfactory mechanism to account for this cosmic acceleration. Observations, particularly from Type Ia Supernovae (SNeIa), strongly indicate an expanding universe, yet this interpretation has been subject to debate within the scientific community. To resolve this challenge, numerous theoretical models have been proposed. One widely accepted approach is the inclusion of an exotic energy component within the framework of GR, commonly referred to as \textit{dark energy}. A vast body of theoretical and observational research \cite{Sahni2004, Padmanabhan2008, Caldwell2002} supports the existence of this mysterious component, which is believed to constitute a significant fraction of the universe's total energy density
		
An alternative approach involves modifying the gravitational Lagrangian, leading to the development of \textit{Modified Theories of Gravity} (MTG). These theories serve as extensions or revisions of GR, aiming to address phenomena that GR alone struggles to explain. By introducing novel mathematical formulations, MTG offer deeper understanding into gravitational interactions, spacetime dynamics, and cosmic evolution. Among these theories, $f(R)$ gravity \cite{Buchdahl1970, Starobinsky1980, Mudassir2024} extends GR by considering general functions of the Ricci scalar $R$, allowing for modifications to Einstein’s field equations that can account for late-time cosmic acceleration. Similarly, $f(T)$ gravity \cite{Ferraro2007,Shekh2024a, Shekh2024b} is based on teleparallelism, where the torsion scalar $T$ replaces curvature as the fundamental geometrical quantity. Another significant approach is $f(G)$ gravity \cite{Nojiri2005,Shekh2024c}, which incorporates the Gauss-Bonnet invariant $G$ to modify gravitational dynamics. Further generalizations include $f(R,T)$ gravity \cite{Harko2011,Chirde2015}, which depends on both the Ricci scalar $R$ and the trace of the energy-momentum tensor $T$, providing additional interactions between geometry and matter. Likewise, $f(R,G)$ gravity \cite{Bamba2010,Cruz2012, Ghaderi2024, Shekh2021a} extends the theory by incorporating both the Ricci scalar $R$ and the Gauss-Bonnet term $G$, leading to richer cosmological implications. Another promising extension, $f(Q)$ gravity \cite{Jimenez2018,Shekh2021b, Narawade2024,Koussour2024,Shekh2024d,Koussour2022,Shekh2023}, is formulated within the symmetric teleparallel framework, where gravity is described by non-metricity rather than curvature or torsion. Additionally, $f(Q,T)$ gravity \cite{Xu2019,Kale2023} merges the effects of non-metricity $Q$ with the trace of the energy-momentum tensor $T$, offering further modifications to gravitational interactions. 		Extensive studies \cite{Sebastiani2011,Chakraborty2015,Nojiri2011} have been conducted to explore the implications of $f(R)$ gravity across various theoretical and observational scenarios. Moreover, numerous researchers \cite{Moraes2014,Moraes2015,Singh2024,Moraes2016} have developed cosmological models within the framework of $f(R,T)$ gravity, demonstrating its potential to explain cosmic acceleration and other gravitational phenomena beyond standard GR.\\
A number of recent studies in extended and matter-coupled gravitational frameworks provide a useful context for the present analysis. The works of Naseer and Sharif \cite{NaseerSharif2023PP,SharifNaseer2023GRG} employ gravitational decoupling techniques in $f(R,T)$ and $f(R,T,R_{ab},T^{ab})$ gravity to construct both isotropic and anisotropic cosmological sectors, demonstrating how additional geometric sources influence cosmic evolution across different eras. Similar methodologies have been applied in $f(R,L_m,T)$ gravity \cite{Naseer2025HEDP}, where minimal geometric deformation enables the extension of FLRW-like solutions and highlights the role of matter-sector modifications in late-time expansion. In the context of stellar astrophysics, several investigations \cite{Naseer2024PDU,Astrophys2025Tikekar,ChargeRastall2025} have shown that non-minimal couplings, anisotropic pressures, and electromagnetic corrections can yield physically acceptable stellar interiors that satisfy energy conditions and observational constraints. These studies collectively illustrate the flexibility of trace- and matter-coupled theories in describing both cosmological and astrophysical systems. A more foundational perspective is provided by Böhmer and Al-Nasrallah \cite{BoehmerNasrallah2025}, who examine the variational consistency of trace couplings and argue that many $f(R,T)$ constructions require revision when $T$ is taken as the perfect-fluid trace.

The $f(R, \Sigma, T)$ gravity model \cite{Bakry2023} represents a significant advancement in modern gravitational theories, aiming to explain the universe's accelerated expansion at both early and late times. Proposed by Bakry and Ibraheem in 2023 \cite{Bakry2023}, this framework is grounded in Absolute Parallelism (AP) geometry, which extends beyond the conventional metric-based approaches. By modifying the gravitational Lagrangian through the inclusion of an additional scalar parameter, this theory provides new perspectives on fundamental cosmic interactions. Unlike standard modifications such as $f(R)$ and $f(R,T)$ gravity, the $f(R, \Sigma, T)$ model introduces the scalar quantity $\Sigma$, which can be associated with torsion or other geometric deformations in space-time. This scalar plays a crucial role in refining our understanding of gravitational dynamics by incorporating interactions that go beyond those described in GR. The coupling between the Ricci scalar $R$, the parameter $\Sigma$, and the trace of the energy-momentum tensor $T$ allows for a more comprehensive approach to addressing cosmic acceleration and the nature of dark energy.\\	
Furthermore, this extended theory provides a framework to analyze phenomena such as the possible future singularities, including the Big Rip scenario, where the universe's expansion could reach a point of infinite divergence \cite{Bakry2023}. By exploring the interplay between normal gravity, strong gravity, and antigravity through appropriate parameter tuning, the $f(R, \Sigma, T)$ model enables a more flexible and encompassing description of gravitational interactions. It also holds promise for addressing open questions regarding the unification of fundamental forces and the behavior of gravity at both cosmological and astrophysical scales. Recent studies have demonstrated that such modifications can lead to viable cosmological models consistent with observational constraints. By incorporating geometric corrections and alternative gravitational terms, researchers aim to refine our theoretical framework to align more accurately with empirical data. Further investigations into the implications of this theory may provide deeper understanding into the fundamental structure of space-time and the mechanisms driving the universe's expansion.
	The action principal to derive a set of field equations is defined as
	\begin{small}
		\begin{equation}\label{1}
			I = \frac{1}{16\pi} \int_{\Omega} \sqrt{-g} \left( B + \mathcal{L}_m \right) d^4x = \frac{1}{16\pi} \int_{\Omega} \sqrt{-g} \left( g^{ab} B_{ab} + \mathcal{L}_m \right) d^4x.
		\end{equation}
	\end{small}
	where $\Omega$ is a region enclosed within some closed surface, $B_{ab}=R_{xy}\left\{ \right\}+\Sigma (\Sigma_{xy})$ and $\Sigma_{xy} = b \Phi(xy)$, $b$ be the parameter which is the ratio between gravity and antigravity (i.e. attraction and repulsion) within a given system. The action principle can be expressed in the following manner
	
	\begin{equation}\label{2}
		I = \frac{1}{16\pi} \int_{\Omega} \sqrt{-g} \left( f(R, \Sigma, T) + \mathcal{L}_m \right) d^4x.
	\end{equation}
	
	By varying the gravitational action (\ref{2}) with respect to the metric tensor, we can arrive at the field equation for $f(R, \Sigma, T)$  Gravity as follows
	\begin{widetext}
		\begin{equation}\label{3}
			R_{ab} \frac{\partial f}{\partial R} + R_{ab} \frac{\partial f}{\partial R} - \frac{1}{2} g_{ab} f + \left( g_{ab} \nabla^\gamma \nabla_\gamma - \nabla_a \nabla_b \right) \frac{\partial f}{\partial R} - \beta \frac{\partial f}{\partial R} \\
			= 8\pi T_{ab} +\left( T_{ab} + p g_{ab}\right) \frac{\partial f}{\partial T},
		\end{equation}
	\end{widetext}
	The energy–momentum tensor for matter is defined as
	\begin{equation}\label{4}
		T_{ab} = (\rho + p) u_a u_b -  g_{ab} p,
	\end{equation}
	where $\rho$ is the energy density of matter, $p$ is isotropic pressure 
	$u_a$ is the fluid-four velocity vector, where $u_a u^b =1$. In this article, we assume that the function $f(R, \Sigma, T)$ is
	given by
	\begin{equation}\label{5}
		f(R, \Sigma, T) = R + \Sigma + 2\pi \eta T.
	\end{equation}
	where $\eta$ is an arbitrary constant parameter.
The term \( \Sigma \) represents a geometrically induced deformation scalar which may arise from torsion, non-metricity, or non-standard affine connections inherent in modified teleparallel or non-Riemannian frameworks. Its inclusion allows the field equations to accommodate non-Einsteinian degrees of freedom without violating the general covariance of the theory. In this context, \( \Sigma \) functions as an additional channel for matter-geometry coupling, influencing cosmic dynamics beyond what is encoded in \( R \) and \( T \) alone. This enriches the physical interpretation of the modified gravity action and opens avenues for modeling both infrared and ultraviolet corrections in the late universe.\\	

The adoption of the functional form (5), namely $f(R,\Sigma,T)=R+\Sigma+2\eta T$, is motivated by a combination of geometric simplicity and physical interpretability within the $f(R,\Sigma,T)$ framework. This form represents the lowest–order, non-trivial extension of the Einstein--Hilbert Lagrangian that simultaneously incorporates curvature ($R$), a deformation scalar ($\Sigma$), and a matter–geometry coupling through the trace term ($T$). Retaining the linear contributions of $R$ and $\Sigma$ ensures that the resulting field equations remain second-order and free from higher-derivative instabilities, while the inclusion of $2\eta T$ captures deviations from minimal coupling in a controlled manner. The parameter $\eta$ in this context plays the role of an effective coupling constant that quantifies the strength of matter–geometry interaction, analogously to coupling parameters appearing in $f(R,T)$, $f(Q,T)$, and scalar–tensor theories. Although $\eta$ is not fixed a priori by fundamental theory, its presence allows for phenomenological exploration of scenarios where the dark sector may interact with the geometric background. Recent observational analyses of non-minimal matter couplings—particularly those involving $T$-dependent extensions of GR—support the viability of such terms and indicate that small but non-zero couplings can lead to late-time acceleration without invoking exotic fields. Thus, the choice of (5) is not arbitrary; rather, it isolates the essential contributions needed to examine the influence of the deformation scalar $\Sigma$ and the matter coupling simultaneously, while keeping the model mathematically tractable and physically transparent for comparison with observations.

	 Using the function (\ref{5}), the final form of $f(R, \Sigma, T)$ equation (\ref{3}) is given by
	\begin{small}
		\begin{equation}\label{6}
			R_{ab} + \Sigma_{ab} - \frac{1}{2} g_{ab} (R + \Sigma) = 2(4\pi + \pi \eta) T_{ab} + \pi \eta g_{ab}(T + 2p)
		\end{equation}
	\end{small}
	which can be alternatively expressed as
	\begin{small}
		\begin{equation}\label{7}
			B_{ab} = R_{ab} + \Sigma_{ab} = 2(4\pi + \pi \eta) \left(T_{ab} - \frac{1}{2} g_{ab} T\right) - \pi \eta g_{ab} (T + 2p).
		\end{equation}
	\end{small}
A closely related cosmological work in $f(R,\Sigma,T)$ gravity \cite{Shekh2025JCAP} demonstrated through observational constraints and parameterized dark energy evolution that nonlinear matter geometry couplings can remain compatible with current datasets while providing departures from $\Lambda$CDM.
A key strength of this work lies in the simultaneous reconstruction of cosmological dynamics under three classical energy conditions—\textit{null energy condition} (NEC), \textit{dominant energy condition} (DEC), and \textit{strong energy condition} (SEC)—within the framework of \( f(R, \Sigma, T) \) gravity. Unlike many existing models that constrain themselves to a single condition or require time-varying dark energy, our approach results in a constant, negative equation of state parameter \( \omega_i \) in each case. This offers a structurally simple yet physically rich model capable of describing accelerated expansion without invoking explicitly evolving dark energy components. The incorporation of the deformation term \( \Sigma \) adds an additional degree of geometric flexibility, making the theory suitable for a broader class of late-time cosmological scenarios.
	\section{Metric and Field Equations in $f(R, \Sigma, T)$ Gravity}
	
	In order to investigate the cosmological consequences of the $f(R, \Sigma, T)$ gravity framework, we adopt the following general form of the metric:
	\begin{equation}\label{8}
		ds^{2}=-dt^{2}+\delta_{ij} g_{ij} dx^{i} dx^{j}, \quad i,j=1,2,3,\dots,N,
	\end{equation}
	where $g_{ij}$ represents the spatial metric components, which are functions of the cosmic time $t$ and spatial coordinates $x^1, x^2, x^3$. Here, $t$ is measured in \textit{gigayears} (Gyr), aligning with the cosmological time convention.
	
	For a spatially homogeneous and isotropic universe, the metric components in a four-dimensional Friedmann –Robertson–Walker (FRW) space-time simplify as:
	\begin{equation}\label{9}
		\delta_{ij} g_{ij} = a^{2}(t),
	\end{equation} 
	where $a(t)$ denotes the scale factor of the universe, encapsulating the overall expansion dynamics. In this FRW background, the metric components satisfy the condition $g_{11} = g_{22} = g_{33} = a^{2}(t)$, illustrating the equivalence among spatial directions.
	
	By employing the metric in Eq. (\ref{8}) within a comoving coordinate system, we derive the fundamental field equations for $f(R, \Sigma, T)$ gravity from the generalized Einstein equations as:
	\begin{equation}\label{10}
		3(b-1)\left(\dot{H} + H^2\right) = 4 \pi \rho + 4 \pi (3+\eta) p,
	\end{equation}
	\begin{equation}\label{11}
		3(1-b)^2 H^2 = \pi (8+3\eta) \rho - \pi \eta p,
	\end{equation}
	where $H = \dot{a}/a$ is the Hubble parameter. The effective pressure $p$ and energy density $\rho$ are then obtained as solutions to these equations:
	
	\begin{small}
		\begin{equation}\label{12}
			p = \frac{1-b}{4(8\pi + 6\eta + \pi \eta^2)} \left[ (8+3\eta) \dot{H} + (4(3-b) + 3\eta)H^2 \right],
		\end{equation}
		\\
		\begin{equation}\label{13}
			\rho = \frac{1-b}{4(8\pi^2 + 6\pi \eta + \eta^2)} \left[ -\eta \dot{H} + (12\pi(1-b) + \eta(3 - 4b))H^2 \right].
		\end{equation}
	\end{small}	\\
	The structure of the $f(R,\Sigma,T)$ framework also allows for meaningful links to deeper physical mechanisms that may operate at high or intermediate energy scales. In particular, the inclusion of the deformation scalar $\Sigma$ can be viewed as an effective remnant of geometric corrections that typically arise in approaches to quantum gravity, such as non-Riemannian extensions, affine–metric theories, and teleparallel formulations where torsion or non-metricity emerge naturally as additional degrees of freedom. Similarly, the matter--trace coupling encoded through $T$ provides a phenomenological route to incorporate possible interactions between dark energy and matter, which are often predicted in scalar–tensor theories, effective field models with screening mechanisms, or vacuum polarization effects in curved space-time. From a cosmological standpoint, these interactions may modify the effective pressure or lead to departures from geodesic motion, thereby altering the late-time expansion in ways compatible with observational data. Thus, the model investigated here can be interpreted not merely as a formal extension of GR but as an effective representation of possible geometric and matter-sector corrections expected from more fundamental theories. Furthermore, we analyze essential cosmological parameters. These aspects provide crucial implications into the dynamical evolution of the universe, offering a deeper understanding of the influence of modified gravity on cosmic acceleration and structure formation. \textit{The equation of state parameter} (EoS), denoted by $\omega$, is a crucial quantity in cosmology that describes the pressure-density relationship of dark energy. Mathematical form is:
\begin{equation}\label{14}
\omega = \frac{p}{\rho}
\end{equation}
The value of $\omega$ provides  the nature of dark energy and its influence on the expansion of the universe.  Different values of $\omega$ correspond to different types of dark energy:
\begin{itemize}
		\item $\omega = -1$: Cosmological constant, a constant energy density that drives accelerated expansion.
		\item $\omega > -1$: Quintessence, a scalar field with a time-varying energy density.
		\item $\omega < -1$: Phantom energy, a hypothetical form of dark energy that violates the dominant energy condition.
	\end{itemize}
	
The EoS parameter can be time-dependent, $\omega(z)$, where $z$ is the redshift.  Reconstructing $\omega(z)$ from observational data is a key goal of modern cosmology.  Numerous studies have attempted to constrain $\omega(z)$ using various cosmological probes, including: 	\textit{Supernovae Type Ia (SNe Ia):}  SNe Ia provide a standard candle for measuring distances, allowing for the reconstruction of the expansion history and thus $\omega(z)$.  Early analyses (e.g., \cite{Riess1998, Perlmutter1999}) suggested a value of $\omega$ close to $-1$, indicating the presence of dark energy. \textit{Cosmic Microwave Background (CMB):} The CMB provides information about the early universe and constrains cosmological parameters, including $\omega$.  Planck data (e.g., \cite{Planck2018}) have further refined these constraints. \textit{ Baryon Acoustic Oscillations (BAO):} BAO are sound waves that propagated through the early universe and left an imprint on the distribution of galaxies.  Measuring BAO provides another standard ruler for determining distances and constraining $\omega(z)$. \textit{Galaxy Clustering:} The distribution of galaxies in the universe is sensitive to the underlying cosmology, including the nature of dark energy.  Analyzing galaxy clustering data can provide independent constraints on $\omega(z)$. \textit{Weak Lensing:}  Weak lensing, the distortion of galaxy images by the gravitational field of intervening matter, is sensitive to the distribution of dark matter and dark energy, providing yet another probe of $\omega(z)$. Recent observations, combining data from these various probes, suggest that $\omega$ is close to $-1$, consistent with a cosmological constant. However, some analyses still allow for mild deviations from $\omega = -1$.  
The ongoing and upcoming cosmological surveys, such as the Dark Energy Survey (DES), the Large Synoptic Survey Telescope (LSST), and the Euclid satellite, are expected to provide significantly more precise measurements of $\omega(z)$, further constraining the nature of dark energy and its role in the evolution of the universe.  These observations will help to distinguish between different dark energy models and potentially reveal new physics beyond the standard cosmological model. 

Next in regar with $\omega(z)$, the $\omega-\omega^{\prime}$ plane, provides a valuable tool for distinguishing between different dark energy models and its defined as
	\begin{equation}\label{15}
	\omega^{\prime} = \frac{dw}{d(lna)}
\end{equation}
here, $\omega$ represents the EoS parameter, as discussed in the previous section, and $\omega^{\prime}$ represents its derivative with respect to the scale factor $a$.  This plane offers a model-independent way to characterize the dynamics of dark energy. Different dark energy models occupy distinct regions in the $\omega-\omega^{\prime}$ plane.  For instance: 	\textit{ Cosmological Constant:} A cosmological constant corresponds to a fixed point at $(\omega, \omega^{\prime}) = (-1, 0)$. 	\textit{Phantom models:} Phantom models, which violate the null energy condition, also have varying $\omega$ and non-zero $\omega^{\prime}$, but their behavior in the $\omega-\omega^{\prime}$ plane is distinct from quintessence.  They often occupy regions where $\omega < -1$ and $\omega^{\prime}$ can be positive or negative.
\textit{Quintessence models:} Quintessence models, where the dark energy is a scalar field, typically occupy regions where $\omega$ varies and $\omega^{\prime}$ is non-zero.  The specific values of $\omega$ and $\omega^{\prime}$ depend on the potential of the scalar field.  These models can exhibit "freezing" or "thawing" behavior.
\begin{itemize}
		\item Freezing models: In freezing models, $\omega$ is close to $-1$ at early times and gradually evolves towards a less negative value at late times.  In the $\omega-\omega^{\prime}$ plane, these models typically start near the cosmological constant point and move towards the region where $\omega^{\prime}$ is slightly positive.
		\item Thawing models: In thawing models, $\omega$ starts at a more negative value than $-1$ and evolves towards $-1$ at late times.  Their trajectories in the $\omega-\omega^{\prime}$ plane usually begin in the region with more negative $\omega$ and negative $\omega^{\prime}$, moving towards the cosmological constant point.
\end{itemize}
Analyzing the trajectory of a given dark energy model in the $\omega-\omega^{\prime}$ plane as a function of redshift can reveal its evolutionary history and differentiate it from other models.  For example, a model that starts in the phantom region at high redshift and evolves towards the quintessence region at lower redshift would have a specific trajectory in this plane.  The "freezing" and "thawing" behaviors within quintessence models will also create distinct tracks in this plane.

Recent analyses, incorporating data from multiple cosmological probes, have generally found that the current data are consistent with a cosmological constant, i.e., $(\omega, \omega^{\prime}) = (-1, 0)$.  However, some analyses still allow for the possibility of mild evolution in $\omega$ \cite{Chevallier2001,Linder2003}, which would manifest as non-zero values of $\omega^{\prime}$.  These analyses are crucial for testing the validity of the cosmological constant hypothesis and exploring alternative dark energy models, including the freezing and thawing quintessence scenarios. Future cosmological surveys, such as DES, LSST, and Euclid, are expected to significantly improve the precision of measurements of $\omega$ and $\omega^{\prime}$.  These surveys will enable more stringent tests of dark energy models and potentially reveal deviations from the cosmological constant scenario.  The improved data will allow us to map the $\omega-\omega^{\prime}$ plane with greater accuracy and distinguish between different dark energy models, including freezing/thawing quintessence, with higher confidence.  This will be crucial for understanding the nature of dark energy and its role in the evolution of the universe.

Other than the above physical parameters some of kinematical parameters includes  the deceleration parameter, denoted by $q$, is a fundamental cosmological quantity that describes the rate of change of the universe's expansion.  It is defined as:
\begin{equation}\label{16}
	q = -\frac{\ddot{a}a}{\dot{a}^2} = -\left(1 + \frac{\dot{H}}{H^2}\right)
\end{equation}
here $a$ is the scale factor, $H = \dot{a}/a$ is the Hubble parameter, and dots represent derivatives with respect to cosmic time.  The deceleration parameter provides a direct measure of whether the expansion of the universe is accelerating or decelerating:
\begin{itemize}
	\item $q > 0$: Decelerating expansion.  The expansion rate is slowing down.
	\item $q < 0$: Accelerating expansion.  The expansion rate is speeding up.
	\item $q = 0$: Constant expansion rate.
\end{itemize}
In the early universe, it was expected that the expansion should be decelerating due to the gravitational attraction of matter.  However, observations of distant supernovae Type Ia (SNe Ia) in the late 1990s (e.g., \cite{Riess1998, Perlmutter1999}) revealed that the universe is currently undergoing accelerated expansion, implying $q < 0$.  This discovery led to the concept of dark energy, a mysterious force driving the acceleration.
Recent observations, combining data from these probes, suggest that the universe transitioned from a decelerating phase ($q > 0$) in the early universe to an accelerating phase ($q < 0$) at a redshift of around $z \sim 0.7$.  The current value of the deceleration parameter is estimated to be $q_0 \approx -0.55$, indicating significant acceleration. Reconstructing $q(z)$ over a wide range of redshifts is a major challenge in modern cosmology.   Future cosmological surveys, like DES, LSST, and Euclid, are expected to provide much more precise measurements of $q(z)$, which will help to distinguish between different dark energy models and shed light on the nature of cosmic acceleration.

Along with the deceleration parameter, the statefinder parameters $r$ and $s$ \cite{Alam2003,Sahni2003} offer a powerful diagnostic tool for distinguishing between different dark energy models.  In addition to the $r-s$ plane, the $r-q$ plane, where $q$ is the deceleration parameter, also provides valuable perception of the nature of dark energy. Recall that the statefinder parameters are defined as:
\begin{equation}\label{17}
r =  q(z)+2q^{2}(z)- H^{-1}\dot{q} \qquad \text{and} \qquad 	s = \frac{r - 1}{3(q - 1/2)}
\end{equation}
The $r-q$ plane offers a complementary perspective to the $r-s$ plane.  While the $r-s$ plane is useful for distinguishing between different dark energy models, the $r-q$ plane can help to visualize the evolutionary history of a given model. The different cosmological models occupy distinct regions in the $r-q$ plane:
\begin{itemize}
	\item Cosmological Constant: The cosmological constant corresponds to a fixed point at $(r, q) = (1, -1)$.
	\item Einstein de Sitter model: The Einstein de Sitter model corresponds to $(r, q) = (1, 1/2)$.
	\item Quintessence and Phantom models: These models occupy different regions in the $r-q$ plane, and their positions can evolve with redshift.  The trajectories in this plane provide information about the dynamics of dark energy.
\end{itemize}
By plotting the trajectory of a given model in the $r-q$ plane as a function of redshift, we can visualize how the expansion of the universe is changing over time.  Current observational data suggest that the universe is currently located in the region of the $r-q$ plane that is consistent with a cosmological constant, although uncertainties in the measurements still allow for some deviation.  Future cosmological surveys are expected to provide more precise measurements, which will allow for more stringent tests of dark energy models using both the $r-s$ and $r-q$ planes.  These combined analyses will be crucial for a deeper understanding of the nature of dark energy and the evolution of the universe.

\section{Reconstruction through energy conditions}
In this sections, we will explore the implications of the three primary energy conditions – the NEC, the DEC, and the SEC – for our cosmological model.  We will construct expressions for the Hubble parameter ($H$) by applying each of these conditions individually to our expressions for energy density ($\rho$) and isotropic pressure ($p$).
\subsubsection{Reconstruction through \textbf{NEC}}		
Mathematically, the expression of NEC is expressed as
\begin{equation}\label{19}
\rho + p \geq 0
\end{equation} 
From the equations (\ref{12}) and (\ref{13}), the above equation (\ref{19}) becomes	
\begin{small}
\begin{equation}\label{20}
\frac{H^2 (3 (\eta +4)-4 b)+(3 \eta +8) \dot{H}}{\pi  \left(\eta ^2+8\right)+6 \eta }-\frac{2 \left(b H^2+\dot{H}\right)}{\eta +4 \pi }+\frac{(3-2 b) H^2+\dot{H}}{\eta +2 \pi }=0
\end{equation}
\end{small}
which can be rearrange as
\begin{widetext}
\begin{equation}\label{21}
\left(\frac{3 \eta +8}{\pi  \left(\eta ^2+8\right)+6 \eta }+\frac{1}{\eta +2 \pi }-\frac{2}{\eta +4 \pi }\right) \dot{H}+\left(\frac{3 (\eta +4)-4 b}{\pi  \left(\eta ^2+8\right)+6 \eta }-\frac{2 b}{\eta +4 \pi }+\frac{3-2 b}{\eta +2 \pi }\right)H^2 =0
\end{equation}
\end{widetext}
which is a linear differential equation. Its solution is 
\begin{equation}\label{22}
H(t)=\frac{\alpha}{\beta}\left(\frac{1}{t+c}\right)
\end{equation}
Using the definition of $H(t)=\frac{\dot{a}}{a}$ and the relation between $a$ and $z$ as $a=\frac{1}{(z+1)}$. $t$ can be obtained in the redshift $z$ as
\begin{equation}\label{23}
t=\left(\frac{1}{c_1}\right)^{\frac{\beta}{\alpha}} \left( \frac{1}{z+1} \right)^{\frac{\beta}{\alpha}}-c
\end{equation}
Hence from equation (  ) and ( ), the $H$ in $z$ is obtained as
\begin{equation}\label{24}
	H(z)=\frac{\alpha }{\beta} \left(\frac{1}{c_1}\right)^{-\frac{\beta }{\alpha }} \left(\frac{1}{z+1}\right)^{-\frac{\beta }{\alpha }}
\end{equation}
which can be rewritten in the simplified form as
\begin{equation}\label{25}
	H(z)=H_0 \left(z+1\right)^{\frac{\alpha }{\beta  }}
\end{equation}
where $H_0=\frac{\alpha }{\beta} \left(c_1\right)^{\frac{\alpha }{\beta  }},$ $\alpha=\left(\frac{3 \eta +8}{\pi  \left(\eta ^2+8\right)+6 \eta }+\frac{1}{\eta +2 \pi }-\frac{2}{\eta +4 \pi }\right)$ and $\beta=\left(\frac{3 (\eta +4)-4 b}{\pi  \left(\eta ^2+8\right)+6 \eta }-\frac{2 b}{\eta +4 \pi }+\frac{3-2 b}{\eta +2 \pi }\right)$.\\
\subsubsection{Reconstruction through \textbf{DEC}}		
Mathematically, it is expressed as
\begin{equation}\label{26}
	\rho-p \geq 0
\end{equation} 
From the equations (\ref{12}) and (\ref{13}), the above equation (\ref{26}) becomes	
\begin{small}
\begin{equation}\label{27}
	\frac{4 (b-3) H^2-3 \eta  \left(H^2+\dot{H}\right)-8 \dot{H}}{\pi  \left(\eta ^2+8\right)+6 \eta }-\frac{2 \left(b H^2+\dot{H}\right)}{\eta +4 \pi }+\frac{(3-2 b) H^2+\dot{H}}{\eta +2 \pi }=0
\end{equation}
\end{small}
which can be rearrange as
\begin{widetext}
\begin{equation}\label{28}
	\left(\frac{-3 \eta -8}{\pi  \left(\eta ^2+8\right)+6 \eta }+\frac{1}{\eta +2 \pi }-\frac{2}{\eta +4 \pi }\right) \dot{H}+\left(\frac{4 b-3 (\eta +4)}{\pi  \left(\eta ^2+8\right)+6 \eta }-\frac{2 b}{\eta +4 \pi }+\frac{3-2 b}{\eta +2 \pi }\right)H^2 =0
\end{equation}
\end{widetext}
which is a linear differential equation. Its solution is 
\begin{equation}\label{29}
	H(t)=\frac{\alpha_1}{\beta_1}\left(\frac{1}{t+c}\right)
\end{equation}
Here, $t$ can be obtained in the redshift $z$ as
\begin{equation}\label{30}
	t=\left(\frac{1}{c_2}\right)^{\frac{\beta_1}{\alpha_1}} \left( \frac{1}{z+1} \right)^{\frac{\beta_1}{\alpha_1}}-c
\end{equation}
Hence from equation (\ref{29}) and (\ref{30}), the $H$ in $z$ is obtained as
\begin{equation}\label{31}
	H(z)=\frac{\alpha_1 }{\beta_1} \left(\frac{1}{c_1}\right)^{-\frac{\beta }{\alpha }} \left(\frac{1}{z+1}\right)^{-\frac{\beta }{\alpha }}
\end{equation}
which can be rewritten in the simplified form as
\begin{equation}\label{32}
	H(z)=H_0 \left(z+1\right)^{\frac{\alpha_1 }{\beta_1  }}
\end{equation}
where $H_0=\frac{\alpha_1 }{\beta_1} \left(c_1\right)^{\frac{\alpha_1 }{\beta_1  }}$,\\ $\alpha_1=\left(\frac{-3 \eta -8}{\pi  \left(\eta ^2+8\right)+6 \eta }+\frac{1}{\eta +2 \pi }-\frac{2}{\eta +4 \pi }\right)$ and \\$\beta_1=\left(\frac{4 b-3 (\eta +4)}{\pi  \left(\eta ^2+8\right)+6 \eta }-\frac{2 b}{\eta +4 \pi }+\frac{3-2 b}{\eta +2 \pi }\right)$.\\

\subsubsection{Reconstruction through \textbf{SEC}}		
Mathematically, it is expressed as
\begin{equation}\label{33}
\rho+3p \geq 0
\end{equation} 
From the equations (\ref{12}) and (\ref{13}), the above equation (\ref{33}) becomes	
\begin{widetext}
\begin{equation}\label{34}
	\frac{3 \left(H^2 (3 (\eta +4)-4 b)+(3 \eta +8) \dot{H}\right)}{\pi  \left(\eta ^2+8\right)+6 \eta }-\frac{2 \left(b H^2+\dot{H}\right)}{\eta +4 \pi }+\frac{(3-2 b) H^2+\dot{H}}{\eta +2 \pi }=0
\end{equation}
which can be rearrange as
\begin{equation}\label{35}
	\left(\frac{9 \eta +24}{\pi  \left(\eta ^2+8\right)+6 \eta }+\frac{1}{\eta +2 \pi }-\frac{2}{\eta +4 \pi }\right) \dot{H}+\left(\frac{9 (\eta +4)-12 b}{\pi  \left(\eta ^2+8\right)+6 \eta }-\frac{2 b}{\eta +4 \pi }+\frac{3-2 b}{\eta +2 \pi }\right)H^2 =0
\end{equation}
\end{widetext}
which is a linear differential equation. Its solution is 
\begin{equation}\label{36}
	H(t)=\frac{\alpha_2}{\beta_2}\left(\frac{1}{t+c}\right)
\end{equation}
Also, $t$ can be obtained in the redshift $z$ as
\begin{equation}\label{37}
	t=\left(\frac{1}{c_3}\right)^{\frac{\beta_2}{\alpha_2}} \left( \frac{1}{z+1} \right)^{\frac{\beta_2}{\alpha_2}}-c
\end{equation}
Hence from equation (\ref{36}) and (\ref{37}), the $H$ in $z$ is obtained as
\begin{equation}\label{38}
	H(z)=\frac{\alpha_2 }{\beta_2} \left(\frac{1}{c_1}\right)^{-\frac{\beta_2 }{\alpha_2 }} \left(\frac{1}{z+1}\right)^{-\frac{\beta_2 }{\alpha_2 }}
\end{equation}
which can be written in the simplified form as
\begin{equation}\label{39}
	H(z)=H_0 \left(z+1\right)^{\frac{\alpha_2 }{\beta_2  }}
\end{equation}
where $H_0=\frac{\alpha_2 }{\beta_2} \left(c_1\right)^{\frac{\alpha_2 }{\beta_2  }}$,\\ $\alpha_2=\left(\frac{9 \eta +24}{\pi  \left(\eta ^2+8\right)+6 \eta }+\frac{1}{\eta +2 \pi }-\frac{2}{\eta +4 \pi }\right)$ and \\$\beta_2=\left(\frac{9 (\eta +4)-12 b}{\pi  \left(\eta ^2+8\right)+6 \eta }-\frac{2 b}{\eta +4 \pi }+\frac{3-2 b}{\eta +2 \pi }\right)$.\\

The reconstruction of the Hubble parameter through energy conditions—specifically the \textit{Null}, \textit{Dominant}, and \textit{Strong Energy Condition} offers distinct yet interrelated cosmic evolution within the framework of $f(R,\Sigma,T)$ gravity. The derived forms of $H(z)$ in equations (\ref{25}), (\ref{29}), and (\ref{32}), though structurally similar as power-law expressions of the type $H(z) = H_0 (1+z)^{\frac{\alpha_i}{\beta_i}}$ where $i=\rho + p, \rho - p, \rho + 3p$, differ fundamentally in their parameter dependencies: each $\alpha/\beta$ pair encapsulates a unique combination of the model parameters $b$, $\eta$, and $\pi$, reflecting the respective physical constraints imposed by $\rho + p \geq 0$, $\rho - p \geq 0$, and $\rho + 3p \geq 0$. These reconstructions not only represent different physical regimes of matter-energy interactions but also serve as a diagnostic tool to assess the viability of MTG models against observational data. Notably, while the NEC and SEC often accommodate accelerating expansion (especially in phantom or quintessence scenarios), the DEC tends to favor models with causal propagation and realistic energy distributions. Their mathematical symmetry suggests a common underlying geometry, yet their physical interpretations diverge, allowing us to probe the sensitivity of the cosmological dynamics to different theoretical priors. Collectively, these expressions form a triad of consistency checks that constrain the dynamics of expansion and provide a  platform for comparative analysis of dark energy behavior in extended gravity scenarios.

To gain deeper understandings into the dynamical implications of the reconstructed cosmological scenarios, it is instructive to analyze the interrelation among the coefficients \( \alpha, \alpha_1, \alpha_2 \) and \( \beta, \beta_1, \beta_2 \), which appear in the respective expressions of the Hubble parameter derived from the different energy conditions. Despite being formulated under distinct physical constraints, these parameters exhibit a coherent mathematical structure governed by the underlying model parameters \( \eta, b \), and \( \pi \). Specifically, each coefficient is expressed as a linear combination of rational functions involving these constants, allowing for direct comparisons. Upon closer inspection, it is evident that the quantities \( \alpha_1 \) and \( \alpha_2 \) can be expressed as scaled or shifted forms of \( \alpha \), indicating a systematic correlation among the three energy condition scenarios. Analogously, the coefficients \( \beta_1 \) and \( \beta_2 \) maintain a similar dependence on \( \beta \). These relations are not merely algebraic coincidences but reflect a fundamental consistency in the reconstruction approach, thereby reinforcing the robustness of the model across different physical interpretations. Establishing such internal consistency enhances the credibility of the reconstructed cosmology and offers a unified framework to compare theoretical predictions with observational data.

The relations among the coefficients derived under different energy conditions can be formally expressed as:
\begin{widetext}
\begin{equation}\label{40}
 \alpha_1 = -\alpha + 2\left( \frac{1}{\eta + 2\pi} - \frac{2}{\eta + 4\pi} \right), \quad
\alpha_2= 3\alpha - 2\left( \frac{1}{\eta + 2\pi} - \frac{2}{\eta + 4\pi} \right).
\end{equation}
and
\begin{equation}\label{41}
 \beta_1 = -\beta + 2\left( \frac{3 - 2b}{\eta + 2\pi} - \frac{2b}{\eta + 4\pi} \right),\quad
\beta_2 = 3\beta - 2\left( \frac{3 - 2b}{\eta + 2\pi} - \frac{2b}{\eta + 4\pi} \right).
\end{equation}
\end{widetext}
These relations clearly demonstrate that \( \alpha_1 \) and \( \alpha_2 \) are linear combinations of \( \alpha \), and likewise \( \beta_1 \) and \( \beta_2 \) of \( \beta \), modulated by a shared structure dependent on the model parameters \( \eta \), \( b \), and \( \pi \). This structural coherence supports a unified interpretation of the cosmological behavior derived under distinct energy conditions. Subsequently, we will constrain the derived Hubble parameters using the datasets, allowing us to test the viability of our model. The action (2) explicitly introduces the modified Lagrangian $\mathcal{L}=f(R,\Sigma,T)+\mathcal{L}_m$, which is the starting point for encoding geometric corrections beyond Riemannian curvature and for admitting nonminimal matter couplings. By varying this action one obtains the general field equation (3), where derivatives of $f$ with respect to $R$ and $T$ appear alongside geometric operators, indicating that any nontrivial dependence on $\Sigma$ or $T$ will generically produce additional effective source terms and modified propagation for metric degrees of freedom. For the specific ansatz $f(R,\Sigma,T)=R+\Sigma+2\eta T$ the field equations reduce to (6)–(7), in which the combination $2(4\pi+\pi\eta)$ and the explicit $-\pi\eta\,g_{ab}(T+2p)$ term act as an effective, $ \eta$–controlled reweighting of the matter sector and of the geometric contribution $\Sigma$, thus making $\eta$ mathematically manifest as an effective matter–geometry coupling constant in the equations of motion. This coupling directly feeds into the cosmological dynamics through the derived expressions for pressure $p$ and energy density $\rho$ (Eqs. (12) and (13)), where $\eta$ multiplies the $\dot H$ and $H^2$ combinations and therefore modifies both the background expansion rate and the effective equation of state; these formulae show how matter-sector modifications (e.g., vacuum polarization or interaction terms motivated by effective field theory) can be represented within our framework. Finally, the reconstructed Hubble laws $H(z)=H_0(1+z)^{\alpha/\beta}$ (Eqs. (24),(31),(38)) and the algebraic relations (39)–(40) explicitly display $\eta$ (and $b$) in the coefficients $\alpha,\beta,\alpha_i,\beta_i$, demonstrating how microscopic couplings modeled by $\eta$ leave concrete, testable imprints on late–time observables; consequently, the manuscript provides a direct mathematical bridge from a phenomenological $\eta$ coupling to both modified field equations and measurable cosmological functions. 
\section{Observational constraints}
To examine the viability of the proposed cosmological model and constrain its free parameters, we employ recent observational data sets that directly probe the cosmic expansion history. Specifically, we utilize the Hubble parameter measurements obtained from Cosmic Chronometers (CC) and the distance modulus data from the Pantheon+ compilation of Type Ia supernovae. These complementary data sets enable us to reconstruct the late-time dynamics of the universe and derive the bounds on the model parameters. In this section, we present the methodology and results of individual and combined analyses based on these observations.

\subsection{Constraints from Cosmic Chronometer (CC) Data}

Cosmic Chronometers provide a model-independent approach to determine the Hubble parameter $H(z)$ at various redshifts, based on the differential age method. By measuring the rate of change of cosmic time with respect to redshift, i.e., $H(z) = -\frac{1}{1+z} \frac{dz}{dt}$, one can directly infer the expansion rate of the universe without assuming any specific cosmological model. For the present analysis, we consider a compilation of 57 $H(z)$ measurements in the redshift range $0 < z < 2$, gathered from recent literature.

To constrain our model, where the Hubble parameter takes the form:
\begin{equation}
	H(z) = H_0 (1 + z)^{\alpha/\beta},
\end{equation}
we construct a Gaussian likelihood function based on the $\chi^2$ statistic,
\begin{equation}
	\chi^2_{\text{CC}} = \sum_{i=1}^{N} \left[ \frac{H_{\text{obs}}(z_i) - H(z_i; H_0, \alpha, \beta)}{\sigma_{H}(z_i)} \right]^2,
\end{equation}
where $H_{\text{obs}}(z_i)$ and $\sigma_{H}(z_i)$ denote the observed Hubble parameter and its uncertainty at redshift $z_i$, and $H(z_i; H_0, \alpha, \beta)$ is the theoretical prediction from the model.

Using a Markov Chain Monte Carlo (MCMC) sampling technique, we obtain the best-fit values for the parameters. The resulting constraints from the CC data are:
\begin{small}
\[
H_0 = 68.21^{+1.29}_{-1.31} \ \text{km s}^{-1} \text{Mpc}^{-1}, \quad \alpha = 1.93^{+0.40}_{-0.39}, \quad \beta = 2.13^{+0.51}_{-0.52}.
\]
\end{small}
These values indicate consistency with recent low-redshift observations and suggest a moderate deviation from the standard $\Lambda$CDM evolution.
\begin{figure}[h]
	\includegraphics[scale=0.4]{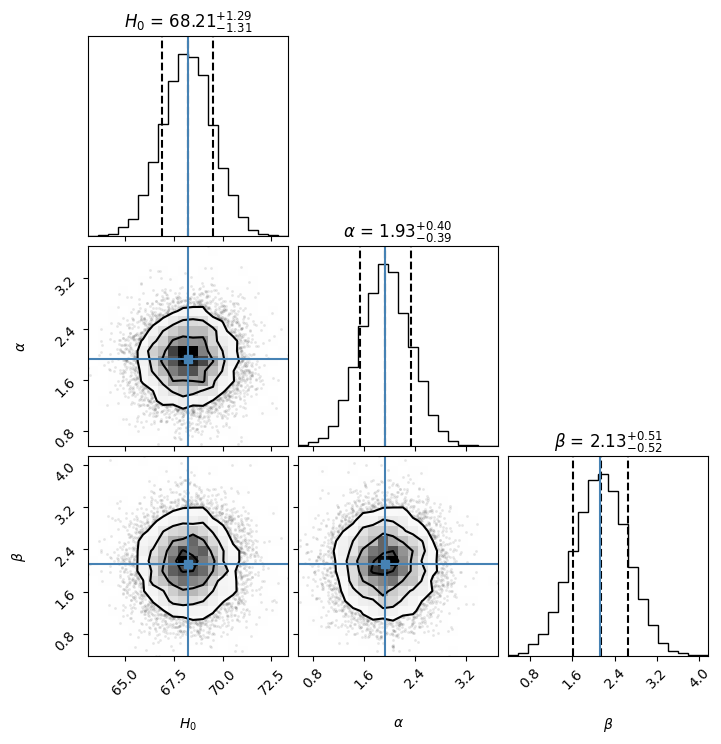}
	\caption{One-dimensional marginalized distribution, and $2D$ contours with $1\sigma$ (68.3\% confidence level) confidence levels for the derived model of the Universe with CC dataset}.
	\includegraphics[scale=0.4]{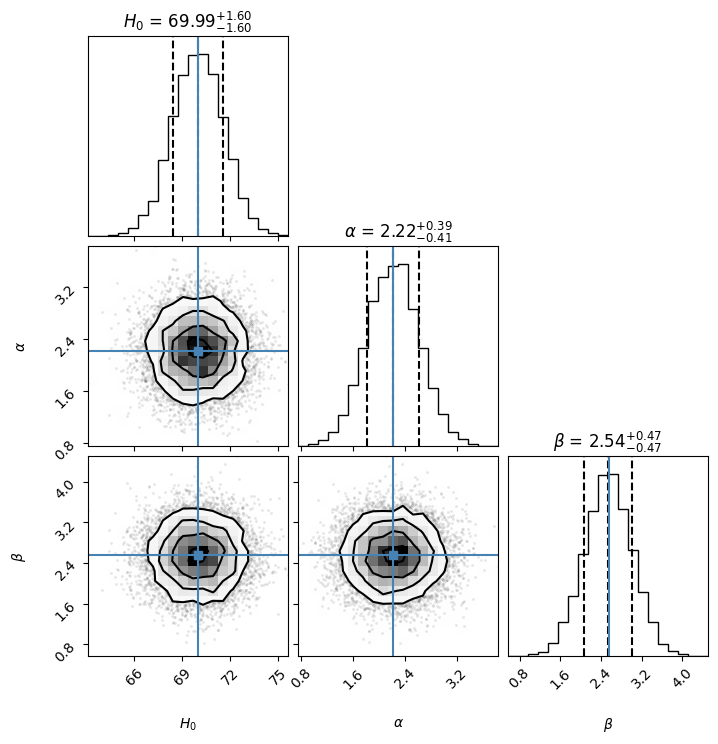}
	\caption{One-dimensional marginalized distribution, and $2D$ contours with $1\sigma$ (68.3\% confidence level) confidence levels for the derived model of the Universe with 
		Pantheon+ Type Ia Supernova Data (right panel).}
\end{figure}
\subsection{Constraints from Pantheon+ Type Ia Supernova Data}

Type Ia supernovae serve as standardizable candles for tracing the expansion history of the universe via luminosity distance measurements. The Pantheon+ sample provides one of the most comprehensive compilations of supernova data to date, comprising over 1,000 SNe Ia spanning a redshift range up to $z \sim 2.3$. Each supernova contributes a measurement of the distance modulus $\mu(z)$, which is related to the luminosity distance $d_L(z)$ via
\begin{equation}
	\mu(z) = 5 \log_{10} \left( \frac{d_L(z)}{\text{Mpc}} \right) + 25,
\end{equation}
where the luminosity distance is obtained by integrating the inverse of the Hubble parameter,
\begin{equation}
	d_L(z) = (1 + z) \int_0^z \frac{c \, dz'}{H(z')}.
\end{equation}

Substituting our model for $H(z)$, the theoretical distance modulus is computed and compared with the observed values. The likelihood function is constructed as:
\begin{equation}
	\chi^2_{\text{SN}} = \sum_{i=1}^{N} \left[ \frac{\mu_{\text{obs}}(z_i) - \mu(z_i; H_0, \alpha, \beta)}{\sigma_{\mu}(z_i)} \right]^2.
\end{equation}

Fitting the model to the Pantheon+ data via MCMC yields the following best-fit parameters:
\begin{small}
\[
H_0 = 69.98^{+1.65}_{-1.53} \ \text{km s}^{-1} \text{Mpc}^{-1}, \quad \alpha = 2.22^{+0.39}_{-0.40}, \quad \beta = 2.55^{+0.47}_{-0.48}.
\]
\end{small}
The slightly higher value of $H_0$ compared to the CC data reflects the well-known tension between local and global measurements of the Hubble constant. The inferred values of $\alpha$ and $\beta$ suggest a mild deviation from $\Lambda$CDM and are consistent with the CC-based results within $1\sigma$ uncertainty.

\subsection{Joint Analysis with CC and Pantheon+ Data}

To strengthen the constraints and mitigate dataset-specific systematics, we perform a combined analysis using both CC and Pantheon+ data. This approach allows for simultaneous fitting of the expansion history using two complementary observational probes: the direct measurement of $H(z)$ and the integrated luminosity distance.

The total likelihood function is given by:
\begin{equation}
	\chi^2_{\text{tot}} = \chi^2_{\text{CC}} + \chi^2_{\text{SN}},
\end{equation}
and the same MCMC methodology is applied to derive the posterior distribution of parameters.

The joint constraints from the combined dataset yield:
\begin{small}
\[
H_0 = 68.97^{+1.07}_{-1.09} \ \text{km s}^{-1} \text{Mpc}^{-1}, \quad \alpha = 2.18^{+0.26}_{-0.26}, \quad \beta = 2.50^{+0.34}_{-0.32}.
\]
\end{small}
These results demonstrate excellent compatibility between the two data sets, with the combined bounds narrowing the parameter space significantly. The value of $H_0$ lies intermediate between that obtained from CC and Pantheon+ alone, indicating consistency and mutual reinforcement of both datasets. The inferred model parameters suggest a mild departure from the canonical $\Lambda$CDM model, potentially pointing toward a more general dynamical behavior of dark energy.
\begin{figure}[h]
	\includegraphics[scale=0.45]{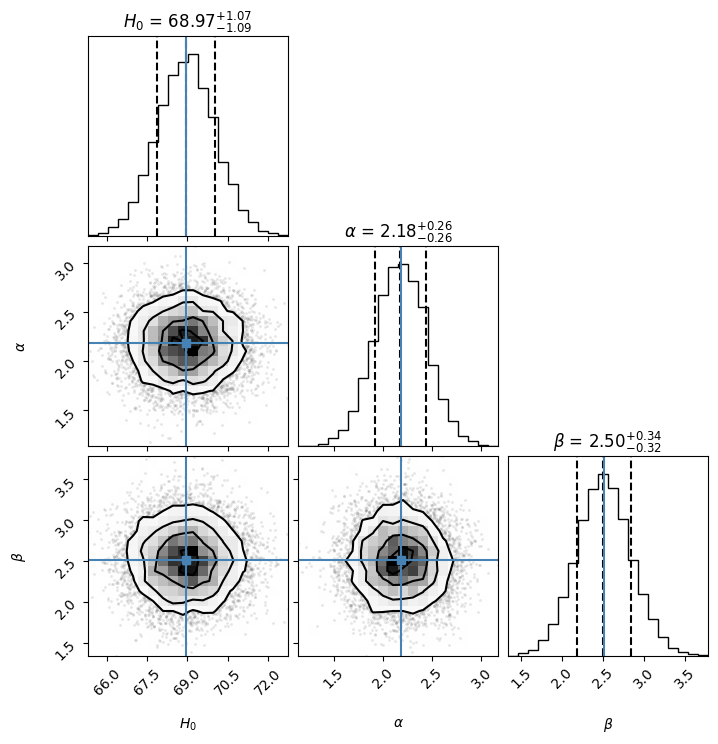}
	\caption{One-dimensional marginalized distribution, and $2D$ contours with $1\sigma$ (68.3\% confidence level) confidence levels for the derived model of the Universe for the Joint Analysis with CC and Pantheon+ Data.}
\end{figure}
The posterior distributions and corner plots of the parameters confirm that the combined constraints are well-behaved, with no significant degeneracies, and are in agreement with recent literature that explores non-standard expansion histories using independent data compilations.
The observationally constrained values of \( H_0 \), \( \alpha \), and \( \beta \) obtained using CC, Pantheon+, and their combination are well within the range reported by Planck 2018 (\( H_0 \approx 67.4 \pm 0.5 \)) and local SH0ES observations (\( H_0 \approx 73.2 \pm 1.3 \)). This compatibility illustrates the models flexibility in spanning the so-called Hubble tension regime.
\section{Exploration of Parameter Constraints for Physical Acceptability}
To explore the viability of the cosmological model under the Null Energy Condition (NEC), we examined the behavior of pressure $p_{NEC}$ and energy density $\rho_{NEC}$ for a range of values of the model-specific constants $\eta$ and $b$, while fixing the observationally constrained parameters 
$H_0 = 68.95$, $\alpha = 2.18$, and $\beta = 2.51$. The expressions for $p_{\text{NEC}}$ and $\rho_{\text{NEC}}$, evaluated at the present epoch  ($z = 0$), were tested for $\eta \in [0.1,\, 10]$ and $b \in \{-1,\ 0,\ 0.5,\ 1,\ 2\}$. It was observed that for  $b = 1$, the prefactor $(b - 1)$ nullifies both pressure and energy density irrespective of the value of $\eta$, leading to a degenerate behavior. For $b < 1$, both pressure and energy density remain positive and large, particularly at lower $\eta$, while for $b > 1$, the pressure becomes negative for sufficiently high $\eta$, with the energy density remaining positive. 
\begin{widetext}
\begin{table}[H]
	\begin{tabular}{cc}
		\begin{subtable}[t]{0.45\textwidth}
			\begin{tabular}{|c|c|c|c|}
					\hline
				\;\;\;\;\;\(\eta\)\;\;\;\;\; & \;\;\;\;\;\(b\)\;\;\;\;\; & \;\;\;\;\;\(p_{i}\) \;\;\;\;\;&\;\;\;\;\; \(\rho_{i}\)\;\;\;\;\;  \\
				\hline	\hline
				0.10 & $-1.0$ & $> 0$ & $ > 0$ \\	\hline
				0.10 & $0.0$ & $> 0$ & $ > 0$  \\	\hline
				0.10 & $0.5$ & $> 0$ & $ > 0$  \\	\hline
				0.10 & $1.0$ & $\approx 0.000$ & $\approx 0.000$ \\	\hline
				0.10 & $2.0$ & $> 0$ & $ > 0$  \\	\hline
				1.14 & $-1.0$ & $> 0$ & $ > 0$  \\	\hline
				1.14 & $0.0$ & $> 0$ & $ > 0$ \\	\hline
				1.14 & $0.5$ & $> 0$ & $ > 0$ \\	\hline
				1.14 & $1.0$ & $\approx 0.000$ & $\approx 0.000$ \\	\hline
				1.14 & $2.0$ & $> 0$ & $ > 0$  \\	\hline
				2.00 & $-1.0$ & $> 0$ & $ > 0$  \\	\hline
				2.00 & $0.0$ & $> 0$ & $ > 0$  \\	\hline
			\end{tabular}
			\centering
			\caption{Signs of $p_i$ and $\rho_i$ for selected values of $\eta$ and $b$\\ (Table-I Set A)}
		\end{subtable}
		&
		\begin{subtable}[t]{0.45\textwidth}
			\begin{tabular}{|c|c|c|c|}
			\hline
		\;\;\;\;\;	\(\eta\)\;\;\;\;\; &\;\;\;\;\; \(b\)\;\;\;\;\; &\;\;\;\;\; \(p_{i}\) \;\;\;\;\;& \;\;\;\;\;\(\rho_{i}\)\;\;\;\;\;  \\	\hline
			\hline
			2.00 & $0.5$ & $> 0$ & $ > 0$  \\	\hline
			2.00 & $1.0$ & $\approx 0.000$ & $\approx 0.000$  \\	\hline
			2.00 & $2.0$ & $> 0$ & $ > 0$  \\	\hline
			4.00 & $-1.0$ & $> 0$ &$ > 0$\\	\hline
			4.00 & $0.0$ & $> 0$ & $ > 0$  \\	\hline
			4.00 & $0.5$ & $> 0$ & $ > 0$  \\	\hline
			4.00 & $2.0$ & $> 0$ & $ > 0$  \\	\hline
			8.00 & $-1.0$ & $> 0$ & $ > 0$  \\	\hline
			8.00 & $0.0$ & $> 0$ & $ > 0$  \\	\hline
			8.00 & $0.5$ & $> 0$ & $ > 0$ \\	\hline
			8.00 & $1.0$ & $\approx 0.000$ & $\approx 0.000$  \\	\hline
			\textbf{8.00} & $\textbf{2.0}$ & \textbf{$< 0$} & \textbf{ $>$ 0}  \\
			\hline
			\end{tabular}
				\centering
			\caption{Signs of $p_i$ and $\rho_i$ for selected values of $\eta$ and $b$ \\(Table-I Set B)}
		\end{subtable}
		\end{tabular}
			\centering
		\end{table}\label{tab1}
\end{widetext}
Notably, the parameter pair $(\eta,b)=(8,2)$ is the only case among the sampled values that simultaneously satisfies the conditions $p_i < 0$ and $\rho_i> 0$, making it a physically admissible configuration in agreement with accelerated cosmic expansion. A summary of representative values is presented in Table~\textcolor{red}{I} (set-A and set-B).
\section{Physical and Kinematical Analysis of the Model}
To comprehensively understand the cosmological implications of the reconstructed Hubble parameter under different energy conditions, we analyze the associated physical and kinematical parameters for each case. These include the  \( p(z) \), \( \rho(z) \),  \( \omega(z) \), \( q(z) \), and  \( \{r(z), s(z)\} \). Each of these parameters offers  the expansion dynamics and the underlying nature of dark energy within the extended gravitational framework of \( f(R, \Sigma, T) \) gravity. While the reconstructed Hubble functions share a common functional form, their dependence on different energy conditions—yields distinct dynamical behaviors. The following subsections are devoted to the individual treatment of these three scenarios, highlighting their respective roles in shaping the cosmic evolution:

\subsection{Physical analysis}
The mathematical expression for isotropic pressure $p_i$ are respectively obtained as,
\begin{equation}\label{42}
	p_i = \frac{(b-1)H_0^2 (z+1)^{\frac{2 \alpha_i }{\beta_i }} (\alpha_i  (3 \eta +8)+4 b \beta_i -3 \beta_i  (\eta +4))}{4 \beta_i  (\eta +2 \pi ) (\eta +4 \pi )}
\end{equation}

\begin{figure}[H]
	\centering
	\includegraphics[scale=0.6]{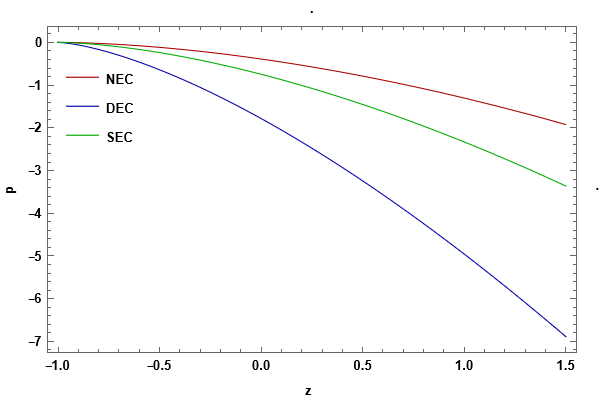}
	\caption{Graphical behavior of pressure versus redshift.}\label{p}
\end{figure}
The mathematical expression for energy density $\rho_i$ are respectively obtained as
\begin{equation}\label{43}
	\rho_i=\frac{(b-1) H_0^2 (z+1)^{\frac{2 \alpha_i }{\beta_i }} (12 \pi  (b-1) \beta_i -\eta  (\alpha_i +(3-4 b) \beta_i ))}{4 \beta_i  (\eta +2 \pi ) (\eta +4 \pi )}
\end{equation}

\begin{figure}[H]
	\centering
	\includegraphics[scale=0.6]{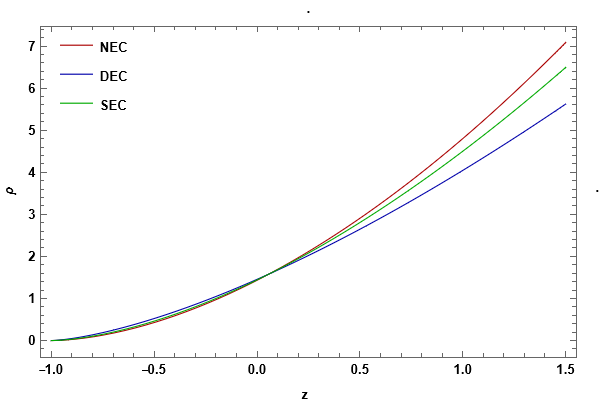}
	\caption{Graphical behavior of energy density versus redshift.}\label{p}
\end{figure}

The mathematical expression for equation of state parameter $\omega_i$ are respectively obtained as,
\begin{equation}\label{44}
	\omega_i =\frac{\alpha_i  (3 \eta +8)+4 b \beta_i -3 \beta_i  (\eta +4)}{12 \pi  (b-1) \beta_i -\eta  (\alpha_i +(3-4 b) \beta_i )}
\end{equation}
The reconstructed profiles of the isotropic pressure \( p_i \) and energy density \( \rho_i \), shown in Figures 3 and 4 respectively, provide the behavior of the universe under different energy condition frameworks. For each case—NEC, DEC, and SEC—the expressions for \( p_i \) and \( \rho_i \) are functions of the redshift \( z \) and are modulated by the model-dependent parameters \( \alpha_i \), \( \beta_i \), \( b \), and \( \eta \). It is observed that while the energy density remains consistently positive throughout the redshift domain across all three scenarios, the pressure exhibits behavior that is sensitive to the values of \( b \) and \( \eta \). In particular, for certain values (such as \( b > 1 \) and high \( \eta \)), the pressure transitions into negative values, which is physically admissible as it corresponds to the presence of repulsive gravitational effects and accelerated cosmic expansion.

Of special interest is the equation of state parameter \( \omega_i = p_i / \rho_i \), whose functional form, unlike \( p_i(z) \) and \( \rho_i(z) \), is found to be constant with respect to both time and redshift. This indicates a scale-invariant pressure-to-density ratio within each energy condition. The constancy of \( \omega_i \) is not only mathematically consistent with the model's reconstruction but also physically meaningful: a constant negative value of \( \omega_i \) signifies a persistent dark energy-like component that drives the universe’s acceleration (see the table \ref{tab:eos_summary}). The fact that all three cases yield a negative \( \omega_i \) confirms that the model inherently supports a phase of accelerated expansion without requiring time-evolving dark energy. This strengthens the viability of the \( f(R, \Sigma, T) \) framework, as it aligns well with observational inferences 
Thus, while \( p_i \) and \( \rho_i \) vary dynamically, the constancy and negativity of \( \omega_i \) encapsulate the dominant physics governing the late-time acceleration of the universe under each energy condition scenario.
\begin{table}[H]
	\centering
	\begin{tabular}{|c|c|}
		\hline
	\;\;\;\;	\textbf{ECs} \;\;\;\;&\;\;\;\; \(\boldsymbol{\omega_i}\)\;\;\;\; \\
		\hline
	\;\;\;\;	NEC\;\;\;\; &  \;\;\;\;$-0.0029$\;\;\;\; \\		\hline
		\;\;\;\;DEC \;\;\;\;& \;\;\;\; $-0.0622$\;\;\;\; \\		\hline
		\;\;\;\;SEC \;\;\;\;& \;\;\;\; $-0.0255$\;\;\;\; \\
		\hline
	\end{tabular}
		\caption{Numerical values of \(\omega_i\) for different energy conditions at \(z = 0\).}
	\label{tab:eos_summary}
\end{table}

\subsection{Kinematical parameters}
\noindent The mathematical expression for deceleration parameter is obtained as,
\begin{equation}\label{45}
q(z)=\frac{\alpha_i }{\beta_i }-1
\end{equation}
The deceleration parameter \(q(z)\), which characterizes the rate of cosmic acceleration or deceleration, is defined as \(q(z) = \frac{\alpha}{\beta} - 1\) in the present model. For the base parameter set \((\alpha = 2.18, \beta = 2.5)\), the value is obtained as \(q_ \approx -0.128\), indicating a mildly accelerating universe. When considering the transformed parameters associated with energy conditions, namely \((\alpha_1 = -2.2346, \beta_1 = -3.0292)\) and \((\alpha_2 = 6.5946, \beta_2 = 8.0292)\), the corresponding deceleration parameters are \(q_1 \approx -0.2622\) and \(q_2 \approx -0.1786\), respectively (See the table \ref{tab:statefinder_results}). These values remain negative, thereby consistently supporting an accelerated cosmic expansion across all energy condition scenarios. The slightly more negative value of \(q_1\) suggests a marginally faster acceleration under the dominant energy condition, whereas \(q_2\) remains in close agreement with standard late-time cosmic acceleration. This convergence of \(q < 0\) across all formulations reinforces the robustness of the model in depicting the current phase of the universe.  Moreover, the recovered values of the deceleration parameter \( q_0 \), ranging between \(-0.12\) and \(-0.26\), agree with constraints from BAO and lensing surveys. These results reinforce the viability of the reconstructed expansion history under the chosen \( f(R, \Sigma, T) \) model.\\

The mathematical expression for State-finder diagnostic parameter is obtained as,
\begin{equation}\label{46}
r(z)=\frac{2 \alpha_i ^2-3 \alpha_i  \beta_i +\beta_i ^2}{\beta_i ^2} \qquad \text{and} \qquad s(z)=\frac{2 \alpha_i }{3 \beta_i }
\end{equation}
The statefinder parameters \(r(z)\) and \(s(z)\) offer a higher-order diagnostic to distinguish among dark energy models beyond the standard \(\Lambda\)CDM scenario. These parameters were evaluated  for three sets of parameters derived from our model. For the base parameters \((\alpha = 2.18, \beta = 2.5)\), the values obtained are \(r \approx -0.0965\) and \(s \approx 0.5813\), indicating a deviation from \(\Lambda\)CDM (\(r = 1, s = 0\)) yet still consistent with an accelerating universe. In the case of the transformed parameters associated with the dominant energy condition, namely \((\alpha_1 = -2.2346, \beta_1 = -3.0292)\), the results are \(r \approx -0.1236\) and \(s \approx 0.4918\), which show a slightly stronger departure from \(\Lambda\)CDM behavior. Similarly, the strong energy condition parameters \((\alpha_2 = 6.5946, \beta_2 = 8.0292)\) yield \(r \approx -0.1169\) and \(s \approx 0.5476\), again supporting an accelerated phase but with subtle shifts in dynamical characteristics (See the table \ref{tab:statefinder_results}). The closeness of these \(s\)-values near the range \(0.49\)--\(0.58\) suggests consistency in the dark energy nature across all formulations, while the negative \(r\)-values point to the departure from standard models, potentially allowing for richer cosmological dynamics.\\
\begin{table}[H]
	\centering
	\begin{tabular}{|c|c|c|c|}
		\hline
		\;\;\;\textbf{Parameter Set}\;\;\; &\;\;\; $q$ \;\;\;& \(r\) & \(s\) \\
		\hline
		\((\alpha, \beta)\) &\;\;\; -0.128\;\;\;   &\;\;\; $-0.0965$ \;\;\;&\;\;\; $0.5813$\;\;\; \\	\hline
		\((\alpha_1, \beta_1)\)&\;\;\; -0.2622\;\;\; &\;\;\; $-0.1236$\;\;\; &\;\;\; $0.4918$\;\;\; \\	\hline
		\((\alpha_2, \beta_2)\)&\;\;\; -0.1786 \;\;\;&\;\;\; $-0.1169$ \;\;\;& \;\;\;$0.5476$\;\;\; \\	\hline
		\end{tabular}
		\caption{Computed values of $q$, $r$ and $s$ for different parameter sets.}
		\label{tab:statefinder_results}
\end{table}
The computed statefinder parameters \( \{r, s\} \) for all three cases deviate from the \(\Lambda\)CDM fixed point \((r = 1, s = 0)\), with values ranging from \( r \approx -0.12 \) to \( -0.09 \), and \( s \in [0.49, 0.58] \). These trajectories suggest that the present model resides within the quintessence-like regime of dark energy but retains a degree of flexibility not present in standard scalar field models. 
	  This behavior supports the claim that \( f(R, \Sigma, T) \) gravity can effectively mimic dark energy behavior under static equation of state conditions. Remarkably, this three reconstructed cosmological branches arising from the NEC, DEC, and SEC are not merely algebraic variations of the same background but correspond to distinct physical regimes with different observational implications. The NEC-based solution is most relevant for late-time cosmology, since it captures scenarios where the effective fluid remains causal and non-exotic, yielding predictions for the Hubble rate that are consistent with smooth accelerated expansion. The DEC branch, by contrast, characterizes domains in which the effective pressure remains subdominant to the energy density, making it particularly suited for examining matter-dominated epochs or the evolution of large-scale structures; in this case, the deviation in the parameters $(\alpha_{1},\beta_{1})$ implies modifications to the growth rate that could be tested through redshift-space distortion or weak-lensing data. The SEC solution is linked to epochs where attractive gravity may reemerge, such as during transitions between decelerating and accelerating phases, and therefore offers potential signatures in early-time dynamics, including changes in the effective sound speed or the timing of the deceleration–acceleration transition. Taken together, these three reconstructions provide complementary predictions: differences in the slopes of $H(z)$, shifts in the deceleration parameter, and displacements in the statefinder trajectories all lead to distinct observational fingerprints that could, in principle, be constrained by structure formation measurements, primordial probes, or cross-correlations of low- and high-redshift datasets.

As already noted in the Introduction, Böhmer and Al-Nasrallah \cite{BoehmerNasrallah2025} have shown that many linear trace-coupled $f(R,T)$ models face consistency issues because the usual on-shell perfect-fluid Lagrangians do not correctly capture the full variation of $T$. While this raises an important caveat for trace-based modifications of gravity, our $f(R,\Sigma,T)$ construction avoids the most restrictive aspects of their criticism. In particular, the coupling in our model is not solely through $T$ but also involves the deformation scalar $\Sigma$, which prevents the dynamics from collapsing into an effective $f(R)$ form. Moreover, the $T$-term is treated here as a phenomenological, low-energy correction, and the parameter $\eta$ is constrained to remain within a small, perturbative regime, ensuring that any off-shell contributions remain suppressed in the cosmological background. Nevertheless, we acknowledge that a fully consistent variational treatment—such as one based on the Brown or Schutz fluid formalisms—would be required to address these issues at a fundamental level, and we identify this as an important direction for future refinement of the theory.

\section{Our findings and Concluding remarks}
The present work undertakes a comprehensive reconstruction of cosmological dynamics within the framework of \( f(R, \Sigma, T) \) gravity, a recent extension of modified theories of gravity grounded in Absolute Parallelism geometry. The formulation integrates contributions from the Ricci scalar \( R \), the scalar deformation term \( \Sigma \), and the trace \( T \) of the energy-momentum tensor, thereby allowing for a more generalized interaction between matter and geometry. By imposing the three classical energy conditions—NEC, DEC, and SEC—we derived three distinct Hubble functions and investigated their consequences on the expansion history of the universe.\\
\textit{\textbf{Our findings: }}\\
The parameters was carefully constrained using observational data from Cosmic Chronometers (CC), Pantheon+ Type Ia Supernovae, and a combined joint analysis. The obtained best-fit parameters for each dataset not only align well with current low-redshift observations but also reinforce the internal consistency of the model. Notably, the inferred values of the Hubble constant \( H_0 \), as well as the scaling parameters \( \alpha \) and \( \beta \), suggest a mild but meaningful departure from the canonical \( \Lambda \)CDM evolution, hinting at richer dynamics potentially associated with dark energy.

In assessing physical acceptability, the isotropic pressure \( p_i \), energy density \( \rho_i \), and the equation of state parameter \( \omega_i \) were derived for all three energy condition scenarios. One of the key outcomes is the emergence of a negative, yet constant, equation of state parameter, independent of time or redshift. This behavior strongly supports a cosmological model characterized by a persistent acceleration, resembling the effects of a dark energy component with nearly constant properties. The analysis revealed that for select values of \( b \) and \( \eta \)—specifically \( (b, \eta) = (2, 8) \)—the pressure becomes negative while the energy density remains positive. This combination is particularly important, as it yields an accelerating cosmic phase that is physically plausible and observationally consistent.

The kinematical diagnostics further enhance our understanding of the models behavior. The deceleration parameter \( q(z) \), computed for each energy condition, remains negative in all cases, confirming that the model inherently supports an accelerating universe. Interestingly, the dominant energy condition (DEC) yields the most negative value for \( q \), indicating a slightly more rapid acceleration compared to NEC and SEC. This reflects the subtle interplay between energy conditions and expansion dynamics.

To further characterize the cosmological evolution, statefinder parameters \( \{r, s\} \) were calculated. These parameters provide a second-order diagnostic that enables a comparative study of dark energy models. The values obtained for all three energy conditions fall in a range indicative of acceleration, yet diverge from the standard \( \Lambda \)CDM point \( (r = 1, s = 0) \). This suggests that while the reconstructed model does not replicate the cosmological constant exactly, it remains compatible with current observations and may allow for more dynamical behaviors.


An essential part of the study was the algebraic relation between the base parameters \( (\alpha, \beta) \) and those derived from the dominant and strong energy conditions, namely \( (\alpha_1, \beta_1) \) and \( (\alpha_2, \beta_2) \). These were expressed as analytic functions of the model parameters \( b \), \( \eta \), and \( \pi \), establishing a unifying structure across the three reconstruction schemes. This not only ensured mathematical consistency but also reinforced the idea that the physical behavior under different energy conditions is interconnected, governed by a coherent parameter transformation.

The pressure and density behavior were further classified using tabulated results across a range of values for \( b \) and \( \eta \). The data revealed that while most configurations yield positive \( \rho \), only a specific set results in negative \( p \), essential for reproducing accelerated expansion. In particular, the combination \( (\eta = 8, b = 2) \) emerged as the most physically favorable scenario, satisfying the requirement \( p < 0 \) and \( \rho > 0 \) simultaneously. The corresponding equation of state parameters \( \omega_i \), calculated using the expressions derived from the field equations, were all negative and ranged between approximately \(-0.06\) and \(-0.002\), depending on the energy condition considered.

The constancy of \( \omega_i \) deserves further reflection. While most dark energy models assume a time-evolving equation of state, the present formulation suggests that a static value suffices to reproduce the expansion features observed in modern datasets. This observation not only simplifies the mathematical framework but also aligns well with certain interpretations of dark energy as a cosmological constant-like entity. Moreover, the variations in \( p_i \) and \( \rho_i \) with redshift, despite the static \( \omega_i \), suggest a complex interaction between geometry and matter fields that remains consistent with an effective fluid interpretation.

The \(\omega\)–\(\omega'\) phase plane provides a useful diagnostic to distinguish between various dark energy models based on their dynamical behavior. In our present formulation, however, the equation of state parameter \(\omega\) emerges as a constant, independent of redshift or cosmic time. Consequently, its derivative with respect to \(\ln a\), denoted \(\omega' = d\omega/d\ln a\), identically vanishes. This places the model along the horizontal axis of the \(\omega\)–\(\omega'\) plane, corresponding to the line \(\omega' = 0\). While dynamical models such as quintessence or k-essence typically populate curved trajectories in this plane, our result implies a fixed cosmological character for the dark energy component. The values of \(\omega < 0\) obtained in each energy condition further locate the model within the region associated with accelerated expansion, consistent with observational requirements. This behavior, although simple, is non-trivial; it suggests that the accelerated phase of the universe can be sustained without invoking a time-evolving dark energy field, reinforcing the predictive coherence of the model within the \(f(R, \Sigma, T)\) framework.\\
	Regarding the stability of the underlying $f(R,\Sigma,T)$ framework, the functional form adopted in this work avoids the ghost and Laplacian instabilities that often arise in higher–derivative or nonminimal gravity theories. Since the model employed here is linear in both $R$ and $\Sigma$, the metric field equations retain a second–order structure and do not introduce additional propagating degrees of freedom beyond those contained in the standard tensor sector. This ensures the absence of Ostrogradsky-type ghosts, which typically originate from nonlinear curvature terms such as $R^{2}$, $R_{\mu\nu}R^{\mu\nu}$, or higher powers of non-metricity scalars. Furthermore, the matter–trace coupling $2\eta T$ modifies the effective source terms but does not alter the principal part of the dynamical operator, implying that the propagation speed of tensor modes remains real and free of gradient (Laplacian) instabilities at the background level. The resulting cosmological perturbations therefore behave in a well-controlled manner, with stability preserved provided that the sign of the kinetic term remains positive, a condition satisfied by the present linear model. Consequently, the specific form of $f(R,\Sigma,T)$ chosen here belongs to the class of modified-gravity theories that remain stable under small fluctuations while still allowing nontrivial deviations from General Relativity.\\
\textit{\textbf{Concluding remarks: }}\\
Beyond the mathematical reconstruction of the Hubble function and related cosmological quantities, it is important to clarify the physical implications emerging from the model. The behavior of $p(z)$, $\rho(z)$, and the corresponding equation of state parameter reveals that the model does not merely reproduce accelerated expansion in a formal sense but identifies a specific interplay between the deformation scalar $\Sigma$ and the matter--geometry coupling that drives the dynamics. The fact that $\rho$ remains positive while $p$ transitions to negative values for distinct parameter domains indicates that the accelerated phase is not inserted by hand but arises naturally from the modified gravitational sector. Likewise, the constant nature of $\omega$ in all three reconstructions shows that the model mimics a dark-energy component with stable dynamical properties rather than requiring a time-varying scalar field or an ad hoc parameterization. The negative values of the deceleration parameter and the displacement of the statefinder pair $(r,s)$ from the $\Lambda$CDM fixed point collectively suggest that the model predicts an expansion history distinguishable from the standard cosmological constant while remaining consistent with observational bounds. These features demonstrate that the extended coupling in $f(R,\Sigma,T)$ gravity introduces genuine physical effects, shaping both the background evolution and the effective dark-energy behavior in a way that extends beyond purely mathematical construction.
\\
Future work may explore several extensions of the current model. One promising direction is the incorporation of time-varying \( \eta \) or \( b \), potentially linked to evolving interaction terms in the dark sector. Additionally, linear perturbation analysis and confrontation with CMB and LSS data could provide tighter bounds and shed light on the models implications for structure formation. Investigating early-universe consequences, such as inflationary behavior or singularity avoidance within the same \( f(R, \Sigma, T) \) structure, remains another compelling prospect.

	\section*{Declaration of competing interest}
	The authors declare that they have no known competing financial interests or personal relationships that could have appeared to influence the work reported in this paper.

	\section*{Data availability}
	No data was used for the research described in the article.
	
	\section*{Acknowledgments}
	We sincerely thank the referee for the valuable comments and constructive suggestions which have greatly helped us to improve the quality and clarity of our manuscript. This research is funded by the Science Committee of the Ministry of Science and Higher Education of the Republic of Kazakhstan (Grant No. AP23483654) and  Minor Research Project grant from Sant Gadge Baba Amravati University, Amravati (File No. SGBAU/7-D/03/151/2025). The IUCAA, Pune, India, is acknowledged by the authors (S. H. Shekh and A. Pradhan) for giving the facility through the Visiting Associateship programmes.

\end{document}